\documentclass[11pt,journal,onecolumn,draftcls,english]{IEEEtran}

\usepackage[]{fontenc}
\usepackage[latin1]{inputenc}
\usepackage{verbatim}
\usepackage{amsmath}
\usepackage{graphicx}
\usepackage{amssymb}
\usepackage{color}
\usepackage{ulem}
\usepackage{xspace,bbm,epic,eepic}
\usepackage[noadjust]{cite} 
\usepackage{theorem}

\title{Coherence-based Partial Exact Recovery Condition for OMP/OLS}

\author{C.~Herzet$^\star$,~C.~Soussen,~J.~Idier,~and~R.~Gribonval
 \thanks{C.~Herzet and R.~Gribonval are with INRIA Rennes - Bretagne
    Atlantique, Campus de Beaulieu, F-35042 Rennes Cedex, France
    (e-mail: Cedric.Herzet@inria.fr; Remi.Gribonval@inria.fr).}
 \thanks{C.~Soussen is with the \CRAN (\cran). Campus Sciences, B.P.
    70239, F-54506 Vand{\oe}uvre-l\`es-Nancy, France (e-mail:
    Charles.Soussen@cran.uhp-nancy.fr.)}% <-this % stops a space
  \thanks{J.~Idier is with the Institut de Recherche en Communications
    et Cybernétique de Nantes (IRCCyN, UMR CNRS 6597), BP 92101, 1 rue
    de la No\"e, 44321 Nantes Cedex~3, France (e-mail:
    Jerome.Idier@irccyn.ec-nantes.fr).}
}%\name {XXX and YYY}

\def\XS{\xspace}
\DeclareMathAlphabet{\mathb}{OML}{cmm}{b}{it}
\def\tb{{\sbm{t}}\XS}

\RequirePackage{ifthen}
\RequirePackage{calc}

\newcommand{\taille}[1][\scad]{%
\ifthenelse{#1 = -5}{}{}%
\ifthenelse{#1 = -4}{\tiny}{}%
\ifthenelse{#1 = -3}{\scriptsize}{}%
\ifthenelse{#1 = -2}{\footnotesize}{}%
\ifthenelse{#1 = -1}{\small}{}%
\ifthenelse{#1 = 0}{\normalsize}{}%
\ifthenelse{#1 = 1}{\large}{}%
\ifthenelse{#1 = 2}{\Large}{}%
\ifthenelse{#1 = 3}{\LARGE}{}%
\ifthenelse{#1 = 4}{\huge}{}%
\ifthenelse{#1 = 5}{\Huge}{}}

\newboolean{@serif}
\setboolean{@serif}{true}
\def\scad{-5} % scadefaultsize

\newcounter{taille}
\newcommand{\sca}[2][\scad]{\setcounter{taille}{#1}%
  \ifthenelse{\boolean{@serif}}
  {{\taille[\thetaille]\textsc{#2}}}
  {\setcounter{taille}{\value{taille}-1}{\uppercase{\taille[\thetaille]#2}}}}

\def\eg{\textit{e.g.,}\XS}
\def\etal{\textit{et al.}\XS}

\def\ie{\textit{i.e.,}\XS}

\def\CRAN{Centre de Recherche en Automatique de Nancy\XS}
\newcommand{\cran}[1][\scad]{\sca[#1]{cran, umr 7039},
   Universit\'e de Lorraine, \sca[#1]{cnrs}\XS}

\newcommand{\adresseCRAN}[1][\scad]{Campus Sciences, B.P. 70239, 
F-5   4506 Vand{\oe}uvre-l\`es-Nancy, France\XS}

                \def\stdpth#1{(#1)}
              \def\stdacc#1{\{#1\}}

   \def\stdscal#1{\langle#1\rangle}

\def\spansub#1{{\mathrm{span}}\stdpth{#1}}

\def\spark#1{{\mathrm{spark}}\stdpth{#1}}

\def\a{{\mathbf a}}
\def\b{{\mathbf b}}

\def\x{{\mathbf x}}
\def\y{{\mathbf y}}

\def\v{{\mathbf v}}

\def\r{{\mathbf r}}
\def\u{{\mathbf u}}

\def\ta{{\tilde{\mathbf a}}}
\def\tb{{\tilde{\mathbf b}}}
\def\tc{{\tilde{\mathbf c}}}

\def\xs{\x^\star}
\def\A{{\mathbf A}}
\def\B{{\mathbf B}}

\def\I{{\mathbf I}}
\def\M{{\mathbf M}}
\def\U{{\mathbf U}}
\def\X{{\mathbf X}}
\def\tA{{\tilde{\mathbf A}}}
\def\tB{{\tilde{\mathbf B}}}
\def\tC{{\tilde{\mathbf C}}}

\def\Diag{\Lambda}

\newcommand{\Qc}{\mathcal{Q}}
\newcommand{\Rc}{\mathcal{R}}
\newcommand{\Qcs}{{\mathcal{Q}^\star}}

\def\am{\arg\min}
\def\ama{\arg\max}

\def\vars_w{\sigma^2_n}
\def\vars{\sigma^2}
\def\ie{\textit{i.e.}, }
\def\eg{\textit{e.g.}, }
\def\etal{\textit{et al.} }
\def\spark{\mathrm{spark}}
\def\spa{\mathrm{span}}

\def\proj{\mathbf{P}_\Qc^\bot }

\def\AQp{\A_{\Qc'}}
\def\xQp{\x_{\Qc'}}
\def\ud{\bar{\delta}}
\def\ld{\underline{\delta}}

\usepackage{soul}

\newtheorem{defi}{Definition}

\newtheorem{proposition}{Proposition}
\newtheorem{lemma}{Lemma}
\newtheorem{theorem}{Theorem}

\begin{document}

\maketitle

%\remCS{Working assumptions: we need to assume that $spark(\A)\geqslant
%  j+1$ in order to define $\mu^{OLS}(j)$ and ensure that
%  $\lb(\ell,j)>0$. Can we reformulate Ths. 1 and 3 with the weaker
%  assumption $\A_{\Qc^\star}$ full rank or with no full rankness
%  assumption at all? It seems that $\mu < 1/(k-1)$ ensures that
%  $\tilde{\A}_{\Qc^\star\backslash\Qc}$ and
%  $\tilde{\B}_{\Qc^\star\backslash\Qc}$ are full rank (proof of Th.
%  2).}

\begin{abstract}
We address the exact recovery of the support of a $k$-sparse vector with Orthogonal Matching Pursuit (OMP) and Orthogonal Least Squares (OLS) in a noiseless setting. We consider the scenario where OMP/OLS have selected good atoms during the first $l$ iterations ($l<k$) and derive a new sufficient and worst-case necessary condition for their success in $k$ steps. Our result is based on the coherence $\mu$ of the dictionary and relaxes Tropp's well-known condition $\mu<1/(2k-1)$ to the case where OMP/OLS have a \emph{partial} knowledge of the support.
\end{abstract}


\section{Introduction}

Sparse representations aim at describing a signal as the combination
of a few elementary signals (or atoms) taken from an overcomplete
dictionary $\A$. In particular, in a noiseless setting, one
wishes to find the vector with the smallest number of non-zero
elements, satisfying a set of linear constraints, that
is %. Specifically, one looks for the solution of
\begin{align}
\min \| \x \|_0\quad \mbox{subject to $\A\x=\y$, } \label{eq:SRproblem}
\end{align}
where $\A\in\mathbb{R}^{m\times n}$, $\x\in\mathbb{R}^{n}$, $\y\in\mathbb{R}^{m}$. Problem \eqref{eq:SRproblem} is usually NP-hard \cite{Natarajan1995Sparse}, that is accessing to the solution requires to sweep over all possible supports for $\x$. 

In order to circumvent this bottleneck, suboptimal (but tractable)
algorithms have been proposed in the literature. Among the most
popular approaches, one can mention the procedures based on a
relaxation of the $\ell_0$ pseudo-norm (\eg Basis Pursuit
\cite{Chen_siam99}, FOCUSS \cite{Gorodnitsky_ieeetsp97}) and the
so-called ``greedy pursuit" algorithms, \eg Matching
  Pursuit (MP)~\cite{Mallat_ieeetsp93}, Orthogonal Matching
  Pursuit (OMP)~\cite{Pati_asilomar93}, Orthogonal Least
  Squares (OLS)~\cite{Chen:1950fk, Natarajan1995Sparse}. However, the
suboptimal nature of these algorithms raises the question of their
performance. In particular, if $\y=\A\xs$, under which conditions can
one ensure that a suboptimal algorithm recovers $\xs$ from $\y$?  The
goal of this paper is to provide novel elements of answer to this
question for OMP and OLS.

 OMP has been widely studied in the recent years, including
  worst case~\cite{Tropp2004Greed,Davenport2010Analysis} and
  probabilistic analyses~\cite{Tropp2007Signal}. The existing exact
  recovery analyses of OMP were also adapted to several extensions of
  OMP, namely regularized OMP~\cite{Davenport2010Analysis}, weak
  OMP~\cite{Foucart2011Stability}, and Stagewise OMP~\cite{Donoho12}.
  Although OLS has been known in the literature for a few decades
  (often under different names \cite{Blumensath2007Difference}), exact
  recovery analyses of OLS remain rare for two reasons. First, OLS is
  significantly more time consuming than OMP, therefore discouraging
  the choice of OLS for ``real-time'' applications, like in
  compressive sensing. Secondly, the selection rule of OLS is more
  complex, as the projected atoms are normalized. This makes the OLS
  analysis more tricky. When the dictionary atoms are close to
  orthogonal, OLS and OMP have a similar behavior, as emphasized
  in~\cite{Foucart2011Stability}. On the contrary, for correlated
  dictionary (\eg in inverse problems), their behavior significantly
  differ and OLS may be a better choice~\cite{Soussen2012Sparsev2}.
  The above arguments motivate our analysis of both OMP and OLS
  although in the present paper, our low mutual coherence assumptions
  imply that the correlation between atoms is weak, therefore we do
  not exhibit difference of behavior between OMP and OLS. 
  
In~\cite{Tropp2004Greed}, Tropp provided the first general analysis of
OMP. More specifically, he derived a sufficient and worst-case
necessary condition under which OMP is ensured to recover a $k$-sparse
vector with a given support, in $k$ iterations.  Recently, Soussen
\etal \cite{Soussen2012Sparsev2} showed that Tropp's exact recovery
condition (ERC) is also sufficient and worst-case necessary for OLS.

A possible drawback of Tropp's ERC stands in its cumbersome
evaluation, since it requires to solve a number of linear systems.
Hence, Tropp proposed in \cite{Tropp2004Greed} a stronger sufficient
condition, easier to evaluate, guaranteeing the recovery of \emph{any}
$k$-sparse vector (for any support) by OMP. His condition
reads:
\begin{align}
\mu < \frac{1}{2k-1}, \label{eq:CBcondition}
\end{align}
where $\mu$ is the dictionary coherence, which only involves inner products between the dictionary atoms (see Definition \ref{def:mu} below). Note that \eqref{eq:CBcondition} is also a sufficient condition for OLS since \eqref{eq:CBcondition} implies Tropp's ERC which, in turn, is a sufficient condition for OLS. 
%The analysis performed by Soussen's \etal in \cite{Soussen2012Sparse} shows that \eqref{eq:CBcondition} is also a sufficient condition for the success of OLS. 
On the other hand, Cai\&Wang recently emphasized that \eqref{eq:CBcondition} is a worst-case necessary condition in some sense~\cite{Cai2010Stable}. 

At this point, let us stress that the conditions mentioned
above are \emph{worst-case} necessary, that is, OMP/OLS will fail
\emph{for some} $\y$'s (and some particular dictionaries for
\eqref{eq:CBcondition}) as soon as they are not satisfied. However,
when these conditions are not verified, one can observe in practice
that OMP/OLS often succeed in recovering $\xs$ for many \emph{other}
observation vectors. In this paper, we investigate the case where
\eqref{eq:CBcondition} is not necessarily satisfied, but OMP/OLS
nevertheless select $l$ atoms belonging to the support of $\x^\star$
during the first $l$ iterations. Our work is in the continuity of
\cite{Soussen2012Sparsev2}, in which the authors extended Tropp's
condition to the $l$-th iteration of OMP and OLS. The resulting
  conditions are however rather complex and
  unpractical for numerical evaluation. In
this paper, we derive a simpler (although stronger) condition based on the coherence
of the dictionary. We show that
\begin{align}
\mu < \frac{1}{2k-l-1}, \label{eq:CBconditionPartial}
\end{align}
is sufficient and worst-case necessary (in some sense) for the success of OMP/OLS in $k$ steps when $l$ atoms of the support have been selected during the first $l$ iterations.

\section{Notations}

The following notations will be used in this paper.
$\stdscal{\,.\,,\,.\,}$ refers to the inner product between vectors,
$\|\,.\,\|$ and $\|\,.\,\|_1$ stand for the Euclidean and the $\ell_1$
norms, respectively. $.^\dag$ denotes the pseudo-inverse of a matrix.
For a full rank and undercomplete matrix, we have
$\X^\dag=(\X^T\X)^{-1}\X^T$ where $.^T$ stands for the matrix
transposition. When $\X$ is overcomplete, $\spark (\X)$ denotes the
minimum number of columns from $\X$ that are linearly
dependent~\cite{Donoho2003Optimally}. $\mathbf{1}_{p}$ (resp
$\mathbf{0}_{p}$) denotes the all-one (resp. all-zero) vector of
dimension $p$. The letter $\Qc$ denotes some subset of the column
indices, and $\X_{\Qc}$ is the submatrix of $\X$ gathering the columns
indexed by $\Qc$. For vectors, $\x_\Qc$ denotes the subvector
  of $\x$ indexed by $\Qc$. We will denote the cardinality of $\Qc$
as $\vert \Qc \vert$. We use the same notation to denote the absolute
value of a scalar quantity. Finally,
$\mathbf{P}_\Qc=\X_\Qc\X_\Qc^\dag$ and $\proj=\I-\mathbf{P}_\Qc$
denote the orthogonal projection operators onto $\spansub{\X_\Qc}$ and
$\spansub{\X_\Qc}^{\perp}$, where $\spansub{\X}$ stands for the column
span of $\X$, $\spansub{\X}^{\perp}$ is the orthogonal complement of
$\spansub{\X}$ and $\I$ is the identity matrix whose dimension is
equal to the number of rows in
$\X$. %We will usually use the shorthand notation ``Oxx" to refer to statement valid for both OMP and OLS.

\section{OMP and OLS} 

In this section, we recall the selection rules defining OMP
and OLS. Throughout the paper, we will assume that the
dictionary columns are normalized.

First note that any vector $\x$ satisfying the constraint in
\eqref{eq:SRproblem} must have a support, say $\Qc$, such that
$\r_\Qc\triangleq \proj \y =\mathbf{0}_m$ since $\y$ must belong to
$\spansub{\A_\Qc}$. Hence, problem \eqref{eq:SRproblem} can
equivalently be rephrased as
\begin{align}
\min \vert \Qc \vert \quad \mbox{subject to $\r_\Qc =\mathbf{0}_m$. } \label{eq:SRproblem2}
\end{align}
OMP and OLS can be understood as iterative procedures searching for a solution of \eqref{eq:SRproblem2} by sequentially updating a support estimate as
%adding one new element to the current support as
\begin{align}
\Qc =  \Qc \cup \{j\},
\end{align}
where
\begin{align}\label{eq:atomselection1}
j \in \left\{
\begin{array}{ll}
\ama_i  \vert \stdscal{\a_i,\r_{\Qc}} \vert  & \textrm{for OMP}\\
\am_i \| \r_{\Qc \cup \{i\}}\|& \textrm{for OLS}
\end{array}\right. 
\end{align}
and $\a_i$ is the $i$th column of $\A$. More specifically, OMP/OLS add
one new atom to the support at each iteration: OLS selects the atom
minimizing the norm of the new residual $\r_{\Qc\cup\{i\}}$
whereas OMP picks the atom maximizing the correlation with the current
residual.

In the sequel, we will use a slightly different, equivalent, formulation of \eqref{eq:atomselection1}. Let us define
\begin{align}
\ta_i & \triangleq \proj \a_i, \\
\tb_i & \triangleq \left\{
\begin{array}{ll}
\frac{\ta_i}{\| \ta_i \|} & \mbox{if $\ta_i\neq\mathbf{0}_m$}\\
\mathbf{0}_m & \mbox{otherwise.}
\end{array}\right. \label{eq:defbi}
\end{align}
Hence, $\ta_i$ denotes the projection of $\a_i$ onto $\spansub{\A_\Qc}^\bot$ whereas $\tb_i$ is a normalized version of $\ta_i$. 
 For simplicity, we dropped the dependence of $\ta_i$ and $\tb_i$ on $\Qc$ in our notations. However, when there is a risk of confusion, we will use $\ta_i^{\Qc}$ (resp. $\tb_i^{\Qc}$) instead of $\ta_i$ (resp. $\tb_i$). With these notations, \eqref{eq:atomselection1} can be re-expressed as 
\begin{align}\label{eq:atomselection2}
j \in \left\{
\begin{array}{ll}
\ama_i  \vert \stdscal{\ta_i,\r_{\Qc}} \vert  & \textrm{for OMP}\\
\ama_i \vert \stdscal{\tb_i,\r_{\Qc}} \vert & \textrm{for OLS}.
\end{array}\right.
\end{align}
The equivalence between \eqref{eq:atomselection1} and
\eqref{eq:atomselection2} is straightforward for OMP by noticing that
$\r_\Qc \in \spansub{\A_\Qc}^\bot$. We refer the reader to
\cite{RebolloNeira2002Optimized} for a detailed calculation for OLS.

Throughout the paper, we will use the common acronym Oxx in
  statements that apply to both OMP and OLS. Moreover, we define the
  unifying notation:
\begin{align}
\tc_i \triangleq
  \left\{
    \begin{array}{ll}
      \ta_i & \textrm{for OMP},\\
      \tb_i & \textrm{for OLS}.
    \end{array}
  \right.
  % \label{eq:ci}
\end{align}
Finally, we will use the notations $\tA$, $\tB$ and $\tC$ to refer to the matrices whose columns are made up of the $\ta_i$'s, $\tb_i$'s and $\tc_i$'s, respectively.

\section{Context and Main Result}

Let us assume that $\y$ is a linear combination of $k$ columns of $\A$, that is 
\begin{align}
\y = \A_\Qcs \x_\Qcs \qquad \mbox{with $\vert \Qcs \vert=k$, $\,x_i\neq 0\  \forall i\in \Qcs$}.
\end{align}
The atoms $\a_i$ ($i\in\Qcs$) will be referred to as the
  ``true'' atoms.  We review hereafter different conditions ensuring
the success of Oxx and present our main result. The definition of
``success" that will be used throughout the paper is as follows.
\begin{defi}[Successful recovery] 
Oxx with $\y$ as input succeeds if and only if it selects atoms in $\Qcs$ during the  first $k$ iterations. 
\end{defi}

The notion of successful recovery may be defined in a weaker sense:
Plumbley~\cite[Corollary~4]{Plumbley2007Polar} first pointed
out that there exist problems for which ``delayed recovery'' occurs
after more than $k$ steps. Specifically, Oxx can select some wrong
atoms during the first $k$ iterations but ends up with a larger
support including $\Qc^\star$ with a number of iterations
  slightly greater than $k$. In the noise-free setting (for
  $\y\in\spansub{\A_{\Qcs}}$), all atoms not belonging to
$\Qc^\star$ are then weighted by 0 in the solution vector.
Recently, a delayed recovery analysis of OMP using restricted-isometry
constants was proposed in~\cite{Zhang2011Sparse} and then extended to
the weak OMP algorithm (including OLS) in~\cite{Foucart2011Stability}.
In the present paper, exactly $k$ steps are performed, thus delayed
recovery is considered as a recovery failure.
  
Moreover, we make clear that in special cases where the Oxx selection rule yields multiple solutions including a wrong atom, that is
 \begin{align}\label{eq:equalitySR}
\max_{i\in\Qcs} \vert \stdscal{\tc_i,\r_{\Qc}}  \vert = \max_{i\notin\Qcs} \vert \stdscal{\tc_i,\r_{\Qc}}  \vert,
\end{align}
we consider that Oxx systematically takes a wrong decision. Hence,
situation~\eqref{eq:equalitySR} always leads to a recovery failure. 

The first thoughtful theoretical analysis of OMP is due to Tropp, see
\cite[Theorems 3.1 and 3.10]{Tropp2004Greed}. Tropp provided a
sufficient and worst-case necessary condition for the exact recovery
of any sparse vector with \emph{a given} support $\Qcs$. The
derivation of a similar condition for OLS is more recent and is due to
Soussen \etal in \cite{Soussen2012Sparsev2}. In the latter paper, the
authors carried out a narrow analysis of both OMP and OLS at
any iteration of the algorithm  using specific recovery
  conditions depending not only on $\Qcs$ but also on the current
  support $\Qc$, whereas Tropp's ERC only involves $\Qcs$ and does not
  depend on the iteration. The main result
in~\cite{Soussen2012Sparsev2} reads:

\begin{theorem}[Soussen \etal 's Partial ERC {\cite[Theorem 3]{Soussen2012Sparsev2}}] \label{th:SoussenERC}Assume that $\A_\Qcs$ is full rank and let $\Qc \subset \Qcs$ with $\vert \Qcs\vert=k$, $\vert \Qc\vert=l$.
 If Oxx with $\y\in\spansub{\A_{\Qcs}}$ as input selects atoms in $\Qc$ during the first $l$ iterations, and
 \begin{align}
\max_{i \notin \Qcs} \| \tC_{\Qcs \backslash \Qc}^\dag \tc_i \|_1 < 1, \label{eq:SoussenERC}
\end{align}
then Oxx only selects atoms in $\Qcs \backslash \Qc$ during the
 $k-l$ subsequent iterations. Conversely, if
\eqref{eq:SoussenERC} does not hold, there exists $\y
\in\spansub{\A_\Qcs}$ for which OLS selects $\Qc$ during the first $l$
iterations and then a wrong atom $j\notin\Qcs$ at the $(l+1)$th
iteration.
\end{theorem}

We note that \eqref{eq:SoussenERC}, on its own, does not constitute a worst-case necessary condition for OMP if $ \Qc \neq \emptyset$. More specifically, as shown in \cite{Soussen2012Sparsev2}, some additional ``reachability" hypotheses are required for \eqref{eq:SoussenERC} to be a worst-case necessary condition for OMP. 
% We refer the reader to \cite{Soussen2012Sparsev2} for a precise definition of the notion of  ``reachability". %Theorem \ref{th:SoussenERC} provides sufficient and worst-case necessary condition for Oxx at any iteration of the algorithm.  

Interestingly, when $\Qc=\emptyset$, one recovers Tropp's ERC \cite{Tropp2004Greed}:
\begin{align}
\max_{i\notin \Qcs} \| \A_{\Qcs}^\dag \a_i \|_1 < 1, \label{eq:TroppERC}
\end{align}
which constitutes a sufficient and worst-case necessary condition for \emph{both} OMP and OLS at the very first iteration. 
%In the general case $\Qc\neq \emptyset$, \eqref{eq:SoussenERC} is a weaker condition than \eqref{eq:TroppERC} which ensures an exact support recovery during the last iterations of Oxx. 

One drawback of Tropp's and Soussen \etal 's ERCs stands in their
unpractical evaluation. Indeed, evaluating 
\eqref{eq:SoussenERC}-\eqref{eq:TroppERC} requires to carry out a
pseudo-inverse (and a projection for \eqref{eq:SoussenERC}) operation.
Moreover, support $\Qcs$ is unknown in practice. Hence,
  ensuring that Oxx will recover any $k$-sparse vector requires to
  test whether~\eqref{eq:TroppERC} is met for all possible supports
  $\Qcs$ of cardinality $k$ (resp. to evaluate~\eqref{eq:SoussenERC}
  for all $\Qcs$ and for all $\Qc\subset\Qcs$ of cardinality $l$).

In order to circumvent this problem, stronger conditions, but easier
to evaluate, have been proposed in the literature. We can mainly
distinguish between two types of ``practical" guarantees: the
conditions based on restricted-isometry constants (RIC) and those
based on the coherence of the dictionary (see Definition \ref{def:mu}
below).

The
contributions~\cite{Davenport2010Analysis,Huang2011Recovery,Liu2012Orthogonal,DBLP:journals/corr/abs-1102-4311,Mo2012Remark}
provide RIC-based sufficient conditions for an exact recovery of the
support in $k$ steps by OMP. The most recent and tightest results are
due to Maleh \cite{DBLP:journals/corr/abs-1102-4311} and Mo\&Shen
\cite{Mo2012Remark}. The authors proved that OMP succeeds in $k$ steps
if $\delta_{k+1}<\frac{1}{\sqrt{K}+1}$, where $\delta_{k+1}$ is the
$(k+1)$-RIC of $\A$. In \cite[Theorem 3.2]{Mo2012Remark}, the authors
showed moreover that this condition is almost tight, \ie there exists
a dictionary $\A$ with $\delta_{k+1}=\frac{1}{\sqrt{K}}$ and a
$k$-term representation $\y$ for which OMP 
%of $k$ columns of $\A$, such that OMP with $\y$ as input 
selects a wrong atom at the first iteration. Let us mention that, by
virtue of Theorem \ref{th:SoussenERC}, these results remain valid for
OLS.

On the other hand, Tropp derived in \cite[Corollary
3.6]{Tropp2004Greed} a sufficient condition for OMP, stronger than
\eqref{eq:TroppERC} but only based on the coherence of the dictionary
$\A$.
\begin{defi} \label{def:mu} The mutual coherence $\mu$ of a dictionary $\A$ is  defined as
\begin{align}
\mu = \max_{i \neq j} \vert \stdscal{\a_i,\a_j}\vert.
\end{align}
\end{defi} 
Tropp's condition reads as in \eqref{eq:CBcondition} and ensures that
\eqref{eq:TroppERC} is satisfied. Since \eqref{eq:TroppERC} guarantees
the success of OLS (Theorem \ref{th:SoussenERC} for iteration $l=0$),
\eqref{eq:CBcondition} is also a sufficient condition for OLS.
Moreover, Cai\&Wang recently showed in \cite[Theorem
3.1]{Cai2010Stable} that \eqref{eq:CBcondition} is also worst-case
necessary in the following sense: there exists (at least) one
$k$-sparse vector $\xs$ and one dictionary $\A$ with
$\mu=\frac{1}{2k-1}$ such that Oxx\footnote{and actually, any sparse
  representation algorithm. } cannot recover $\xs$ from
$\y=\A\xs$. %Finally, the work by Soussen's \etal (see Theorem \ref{th:SoussenERC}) implies that \eqref{eq:CBcondition} is also sufficient and worst-case necessary for OLS.
These results are summarized in the following theorem:
\begin{theorem}[$\mu$-based ERC for Oxx {\cite[Corollary 3.6]{Tropp2004Greed}, \cite[Theorem 3.1]{Cai2010Stable}}]\label{th:SCmu}
If \eqref{eq:CBcondition} is satisfied, then Oxx succeeds in
recovering any $k$-term representation.
%support $\Qcs$ with $\vert \Qcs \vert=k$ for any $\x_\Qcs\in
%\mathbb{R}^{k}$. 
Conversely, there exists an instance of dictionary $\A$ and 
a $k$-term representation for which:
%support $\Qcs$, with $\vert \Qcs \vert=k$, such that: 
\emph{(i)} $\mu=\frac{1}{2k-1}$; \emph{(ii)} Oxx selects a wrong atom
%$j \notin \Qcs$ 
at the first iteration.
\end{theorem}

%{\color{blue} Ubiquity of \eqref{eq:CBcondition} in the guarantees of performance of sparse-representation algorithms? (noisy cases, Basis Pursuit).} 

In this paper, we extend the work by Soussen \etal and provide a
coherence-based sufficient and worst-case necessary condition for the
success of Oxx  in $k$  iterations provided that true atoms have
been selected in the first $l$ iterations. Our  main  result
generalizes Theorem \ref{th:SCmu} to the case where $l$ true
atoms
%in $\Qcs$ 
have been selected:

\begin{theorem}[$\mu$-based Partial ERC for Oxx]\label{th:mainth}
  Consider a $k$-term representation $\y
    \in\spansub{\A_\Qcs}$.  Assume that, at iteration $l<k$, Oxx has
  selected $l$ true atoms in $\Qcs$.
%, with $\vert \Qc\vert=l$, $\vert \Qcs\vert=k$. 
If
\begin{align}
\mu < \frac{1}{2k-l-1}, \label{eq:mainBound}
\end{align}
then Oxx exactly recovers $\Qcs$ in $k$ iterations. 

Conversely, there exists a dictionary $\A$ and a $k$-term
  representation $\y$ such that: \emph{(i)} $\mu=\frac{1}{2k-l-1}$;
  \emph{(ii)} Oxx selects true atoms during the first $l$ iterations
  and then a wrong atom at the $(l+1)$th iteration.
% for at least one \addCS{vector
%   $\x_\Qcs\in\mathbb{R}^{k}$.

% supports $\Qc$, $\Qcs$, with $\vert \Qc \vert=l$, $\vert \Qcs
% \vert=k$, $\Qc \subset \Qcs$, such that: \emph{(i)}
% $\mu=\frac{1}{2k-l-1}$; \emph{(ii)} Oxx selects atoms in $\Qc$
% during the first $l$ iterations and then a wrong atom $j \notin
% \Qcs$ at the $(l+1)$th iteration for at least one \addCS{vector
%   $\x_\Qcs\in\mathbb{R}^{k}$.
\end{theorem}

The proof of this theorem is reported to sections \ref{sec:SCOMP},
\ref{sec:SCOLS} and \ref{sec:NC}. More specifically, we show in
section \ref{sec:SCOMP} (resp. section \ref{sec:SCOLS}) that
\eqref{eq:mainBound} is sufficient for the success of OMP (resp. OLS)
during the last $k-l$ iterations. The proof of this sufficient
condition significantly differs for OMP and OLS. The result is shown
for OMP by deriving an upper bound on Soussen~\etal's extended ERC as
a function of the restricted isometry bounds of the projected
dictionary. As for OLS, the proof is based on a connection between
Soussen~\etal's ERC and the mutual coherence of the normalized
projected dictionary $\tB$. Finally, in section \ref{sec:NC} we prove
that \eqref{eq:mainBound} is worst-case necessary for Oxx in the sense
specified in Theorem \ref{th:mainth}. The proof is common to both OMP
and OLS.

\section{Sufficient condition for OMP at iteration $l$}\label{sec:SCOMP}

In this section, we prove the sufficient condition result of Theorem
\ref{th:mainth} for OMP. The result is a direct consequence of Theorem
\ref{th:ubERC} stated below, which provides an upper bound on the
left-hand side of \eqref{eq:SoussenERC} only depending on the
coherence of the dictionary $\A$:
\begin{theorem}\label{th:ubERC}
Let $\Qc \subset \Qcs$, with $\vert \Qc\vert=l$, $\vert \Qcs\vert=k$. If 
\begin{align}
\mu< \frac{1}{k-1}
\end{align}
then
\begin{align}
\max_{i\notin \Qcs} \| \tA_{\Qc^\star \backslash \Qc}^\dag \tilde{\a}_{i} \|_1\leq \frac{(k-l)\mu}{1-(k-1)\mu}. \label{eq:upperboundERC}
\end{align}
\end{theorem}
The sufficient condition for OMP stated in Theorem \ref{th:mainth} then derives from Theorem \ref{th:ubERC}. We see that
\begin{align}\label{eq:suffcondmu}
\frac{(k-l)\mu}{1-(k-1)\mu}<1
\end{align}
implies \eqref{eq:SoussenERC} and is therefore sufficient for the success of OMP in $k$ iterations. Now, \eqref{eq:suffcondmu} is equivalent to  \eqref{eq:mainBound} which proves the result.

Before proving Theorem \ref{th:ubERC}, we need to define some
quantities characterizing the \emph{projected} dictionary $\tA$
appearing in the implementation of OMP (see \eqref{eq:atomselection2})
and state some useful propositions.
In the following definition, we generalize the concept of restricted
isometry property (RIP) \cite{Candes2005Decoding} to projected
dictionaries, under the name projected RIP (P-RIP):
\begin{defi}\label{def:GRIC}
 Dictionary $\A$ satisfies the P-RIP($\ld_{q,l}$,$\ud_{q,l}$) if and only if $\forall \Qc',
  \Qc$ with $\vert \Qc'\vert=q$, $\vert\Qc\vert=l$, $\Qc \cap
  \Qc'=\emptyset$, $\forall \x_{\Qc'}$ we have
\begin{align}
(1-\ld_{q,l}) \| \x_{\Qc'} \|^2 \leq  \| \tA_{\Qc'}^{\Qc} \xQp \|^2 \leq (1+\ud_{q,l}) \| \x_{\Qc'} \|^2.
\end{align}
 %  $\ub(q,l)$ is said to a be a $(q,l)-$upper bound if $\forall \Qc',
%  \Qc$ with $\vert \Qc'\vert=q$, $\vert\Qc\vert=l$, $\Qc \cap
%  \Qc'=\emptyset$, $\forall \x_{\Qc'}$ we have
%  \begin{align}
%    \ub(q,l) \| \x_{\Qc'} \|^2 \geq \| \tA_{\Qc'}^{\Qc} \xQp \|^2.
%  \end{align}
%  Similarly, $\lb(q,l)\geq 0$ is said to a be a $(q,l)-$lower bound
%  if $\forall \Qc', \Qc$ with $\vert \Qc'\vert=q$, $\vert\Qc\vert=l$,
%  $\Qc \cap \Qc'=\emptyset$, $\forall \x_{\Qc'}$, we have
%  \begin{align}
%\lb(q,l) \| \x_{\Qc'} \|^2 \leq \| \tA_{\Qc'}^{\Qc} \xQp \|^2.
%\end{align}
\end{defi}
The definition of the standard (asymmetric)
restricted isometry constants corresponds to the tightest possible
bounds when $l=0$ (see \eg
\cite{Foucart2009Sparsest,Davies2009Restricted}). For $l\geq
  1$, $\ld_{q,l}$ and $\ud_{q,l}$ can be seen as (asymmetric)
\emph{bounds} on the restricted isometry constants of \emph{projected}
dictionaries. Note that $\ud_{q,l}$ is not necessarily non-negative
since the columns of $\tA$ are not normalized
($\|\ta_i^{\Qc}\|\leq 1$). Note also that many well-known
properties of the standard restricted isometry constants (see
\cite[Proposition 3.1]{Needel_ACMels08} for example) remain valid for
$\ld_{q,l}$ and
$\ud_{q,l}$. %The result stated in Theorem \ref{th:mainth} is a
            %consequence of the two following propositions:

The next proposition provides an upper bound on the left-hand side of \eqref{eq:SoussenERC} only depending on $\ld_{q,l}$ and $\ud_{q,l}$:  
\begin{proposition}\label{UBERC} %If $\lb(k-l,l)>0$, then 
Let $\Qc \subset \Qcs$, with $\vert \Qc\vert=l$, $\vert \Qcs\vert=k$. If $\ld_{k-l,l}<1$, then 
\begin{align}
%\max_{j\notin \Qcs} \| \tA_{\Qc^\star \backslash \Qc}^\dag \tilde{\a}_{j} \|_1 < (\leq ?) (k-l)\, \frac{\ub(2,l)-\lb(2,l)}{2\lb(k-l,l)}.\label{eq:UBERC}
\max_{i\notin \Qcs} \| \tA_{\Qc^\star \backslash \Qc}^\dag \tilde{\a}_{i} \|_1 \leq (k-l)\, \frac{\ud_{2,l}+\ld_{2,l}}{2(1-\ld_{k-l,l})}.\label{eq:UBERC}
\end{align}
\end{proposition}
%As a consequence, if the right-hand side of \eqref{eq:UBERC} is upper bounded by one, Soussen \etal's ERC is satisfied. 
The proof of Proposition \ref{UBERC} is reported to Appendix
\ref{sec:SCOMP}. The next proposition provides some possible values
for  $\ld_{q,l}$ and $\ud_{q,l}$ as a function of the coherence of the
dictionary $\A$:

\begin{proposition} \label{prop:linkRICmu} 
If $\mu<1/(l-1)$, then $\A$ satisfies the P-RIP($\ld_{q,l}$,$\ud_{q,l}$) with 
\begin{align}
\ud_{q,l} &= (q-1)\mu \label{eq:relRICmu1},\\
\ld_{q,l}  &= (q-1)\mu +\frac{\mu^2 q l}{1-(l-1)\mu}.
\label{eq:relRICmu2}
\end{align}
\end{proposition}
The proof of this result is reported to Appendix \ref{sec:SCOMP}. We are now ready to prove Theorem \ref{th:ubERC}:\vspace{0.2cm}

\begin{IEEEproof}
\emph{(Theorem \ref{th:ubERC})} 
%The proof consists in
We rewrite the right-hand side of \eqref{eq:UBERC} as a function of
$\mu$.
% by using Proposition \ref{prop:linkRICmu}.  
From Proposition \ref{prop:linkRICmu}, we have that $\A$ satisfies the
P-RIP($\ld_{q,l}$,$\ud_{q,l}$) with constants defined in
\eqref{eq:relRICmu1}-\eqref{eq:relRICmu2} as long as
\begin{align}
\mu< \frac{1}{l-1}.\label{eq:subcondition_th1}
\end{align} 
Now, we have $\mu<1/(k-1)$ by hypothesis, which implies $\mu<1/(l-1)$.
Using \eqref{eq:relRICmu1} and \eqref{eq:relRICmu2}, we
  calculate that:
\begin{align}
\frac{\ud_{2,l}+\ld_{2,l}}{2}&=\mu+\frac{\mu^2l}{1-(l-1)\mu}
= \frac{\mu(\mu+1)}{1-(l-1)\mu},\\
1-\ld_{k-l,l}&=1-(k-l-1)\mu-\frac{\mu^2(k-l)l}{1-(l-1)\mu}\\
&= \frac{1-(k-2)\mu-(k-1)\mu^2}{1-(l-1)\mu}\\
&=\frac{(\mu+1)(1-(k-1)\mu)}{1-(l-1)\mu}.\label{eq:1moinslambda}
\end{align}
Therefore, the ratio in the right-hand side of \eqref{eq:UBERC}
can be rewritten as
\begin{align}
\frac{\ud_{2,l}+\ld_{2,l}}{2(1-\ld_{k-l,l})}
&= \frac{\mu}{1-(k-1)\mu}.\label{eq:ratio_lambda}
\end{align}
According to~\eqref{eq:1moinslambda},
$\mu<1/(k-1)\leq 1/(l-1)$ implies that $1-\ld_{k-l,l}>0$. 
Proposition~\ref{UBERC} combined with~\eqref{eq:ratio_lambda} implies that~\eqref{eq:upperboundERC} 
is met.
% holds as soon as $1-\ld_{k-l,l}>0$. 
% Using \eqref{eq:relRICmu1} and \eqref{eq:relRICmu2}, the latter condition rewrites
% %\begin{align}
% %\frac{1-(k-2)\mu-(k-1)\mu^2}{1-(l-1)\mu}>0.\label{eq:subcondth1_2}
% %\end{align}
% %Since \eqref{eq:subcondition_th1} is assumed to hold, the denominator is positive and \eqref{eq:subcondth1_2} is equivalent to
% \begin{align}
% 1-(k-2)\mu-(k-1)\mu^2=
% (\mu+1)(1-\mu(k-1))>0. \label{eq:subcondth1_3b}
% \end{align}
% This inequality is satisfied since $\mu<1/(k-1)$ by hypothesis.
\vspace{0.3cm}
\end{IEEEproof}

Before concluding this section, let us remark that unlike
Theorem~\ref{th:SoussenERC}, Theorem \ref{th:mainth} does not
(explicitly) require all $(m\times k)$-submatrices $\A_{\Qcs}$ to be
full rank. However, this condition is implicitly enforced by
\eqref{eq:mainBound}. Indeed, as shown in \cite[Lemma~2.3]{Tropp2004Greed},
\begin{align}
\mu< \frac{1}{k-1}
\end{align}
implies that $\A_\Qcs$ is full rank when $\vert \Qcs\vert=k$.
Hence, since $k-1<2k-l-1$, \eqref{eq:mainBound} also implies that any
submatrix $\A_{\Qcs}$ with $\vert \Qcs\vert=k$ is full rank.
Finally, we remark that the full rankness of $\A_{\Qcs}$
  implies that the projected submatrices $\tA_{\Qc^\star \backslash
    \Qc}$ involved in Theorem~\ref{th:ubERC} are also full
  rank~\cite[Corollary~3]{Soussen2012Sparsev2}.

\section{Sufficient condition for OLS at iteration $l$}\label{sec:SCOLS}

We now prove the sufficient condition for OLS stated in
Theorem \ref{th:mainth}. The result is a consequence of Proposition
\ref{lem:connect_OLS} and Lemma \ref{prop:bound_mutilde} stated below.
We first need to introduce the coherence of the \emph{normalized}
projected dictionary $\tilde{\B}$:
%(denoted by $\tilde{\B}$ with $\tilde{\b}_i=\tilde{\a}_i/\|\tilde{\a}_i\|$ if $i\notin\Qc$, and \zerob otherwise). 
%
%
%
\begin{defi}[Coherence of the normalized projected
  dictionary]\label{def:coherencePD2}
\begin{align}
\mu^{OLS}_l = \max_{\vert \Qc \vert=l}\max_{i \neq j}  \vert \langle 
\tilde{\b}_i^{\Qc}, \tilde{\b}_j^{\Qc} \rangle\vert. 
\end{align}
\end{defi}
%
%The following two lemmas aim to upper bound $\mu^{OLS}_l$ and the
%term $ \max_{i\notin \Qc^\star \backslash \Qc} \| \tilde{\B}_{\Qc^\star \backslash \Qc}^\dag \tilde{\b}_{i}
%\|_1$ involved in \eqref{eq:SoussenERC}.
%

The following proposition gives a sufficient condition on $\mu^{OLS}_l$ under which \eqref{eq:SoussenERC} is satisfied:
\begin{proposition} \label{lem:connect_OLS}
%
%Let $\Qc\subset \Qcs$ with $\vert \Qc \vert=l$, $\vert \Qcs \vert=k$. 
  Let $\Qc \subset \Qcs$, with $\vert \Qc\vert=l$, $\vert
  \Qcs\vert=k$. Assume that $\A_{\Qcs}$ is full rank. If
  $\mu^{OLS}_l<1/(2k-2l-1)$,
%  If 
%\begin{align}
%\mu^{OLS}_l<1/(2k-2l-1), 
%\end{align}
 then 
 \begin{align}\label{eq:ERCOLS}
\max_{i\notin \Qc^\star} \| \tilde{\B}_{\Qc^\star
    \backslash \Qc}^\dag \tilde{\b}_{i} \|_1<1. 
\end{align}
\end{proposition}
\begin{IEEEproof}
When $\tilde{\b}_{i}=\mathbf{0}$, the result is obvious.
    When $\tilde{\b}_{i}\ne\mathbf{0}$, apply~\cite[Corollary
  3.6]{Tropp2004Greed} (that is: if $\A$ has normalized
  columns and $\mu<1/(2k-1)$ then Tropp's ERC is satisfied, \ie
  $\forall \Qc^\star$ such that $\vert \Qc^\star\vert=k$,
  $\max_{i\notin \Qcs} \|\A_{\Qc^\star}^\dag {\a}_{i} \|_1<1$) to the
  matrix $\tilde{\B}$ and to ${\Qc^\star\backslash\Qc}$ of size $k-l$.
  The atoms of $\tilde{\B}_{\Qc^\star\backslash\Qc}$ are of
    unit norm (actually, $\tilde{\B}_{\Qc^\star\backslash\Qc}$ is full
    rank) because $\A_{\Qcs}$ is full
    rank~\cite[Corollary~3]{Soussen2012Sparsev2}. 
\end{IEEEproof}\vspace{0.4cm}

The next lemma provides a useful upper bound on $\mu^{OLS}_l$ as a function of the coherence $\mu$ of the dictionary $\A$:
\begin{lemma} \label{prop:bound_mutilde}
If $\mu<1/l$, then
\begin{align}
  \mu^{OLS}_l\leqslant \frac{\mu}{1-l\mu}. \label{eq:bound_mutilde}
\end{align}
\end{lemma}

The proof of this result is reported to Appendix \ref{annex:SCOLS}.
The sufficient condition stated in Theorem \ref{th:mainth} for OLS
then follows from the combination of Proposition \ref{lem:connect_OLS}
and Lemma \ref{prop:bound_mutilde}. Indeed, \eqref{eq:mainBound}
implies $\mu<1/(k-1)\leq 1/l$ since $2k-l-1=k-1+(k-l)> k-1\geq
  l$. Hence, the result follows by first applying
Lemma~\ref{prop:bound_mutilde}:
  \begin{align}
  \mu^{OLS}_l&\leqslant \frac{\mu}{1-l\mu} < \frac{1}{2k-2l-1},
\end{align}
and then Proposition~\ref{lem:connect_OLS}, which implies that
\eqref{eq:ERCOLS} is met. $\mu<1/(k-1)$ implies that the full
  rank assumption of Proposition~\ref{lem:connect_OLS} is met for any
  $\Qcs$ of cardinality $k$~\cite[Lemma~2.3]{Tropp2004Greed}.

%\begin{theorem}[Partial Sufficient ERC for OLS]\label{th:mainth_ols}
%  Assume that, at the $j$th iteration, OLS has selected atoms in $\Qc
%  \subset \Qcs$, with $\vert \Qc\vert=j$, $\vert \Qcs\vert=k$. If
%  \begin{align}
%    \mu < \frac{1}{2k-j-1}, \label{eq:mainBound_copy}
%  \end{align}
%  then OLS exactly recovers the support of the sparse vector. 
%\end{theorem}
%
%\begin{IEEEproof}
%  \eqref{eq:mainBound_copy} implies that $\mu<1/j$ since
%  $2k-j-1=k+(k-j-1)\geqslant k\geqslant j$.  Apply
%  Lemma~\ref{prop:bound_mutilde}:
%  \begin{align*}
%  \mu^{OLS}_j&\leqslant \frac{\mu}{1-j\mu}{\color{red}<} \frac{1}{2k-2j-1}
%\end{align*}
%and then Lemma~\ref{lem:connect_OLS}: the ERC-OLS($\A,\Qc^\star,\Qc$)
%is met.
%\end{IEEEproof}

\section{Worst-case necessary condition for Oxx at iteration $l$}\label{sec:NC}

Cai\&Wang recently showed in \cite[Theorem 3.1]{Cai2010Stable} that
there exist dictionaries $\A$ with $\mu = \frac{1}{2k-1}$ and linear
combinations $\y$ of $k$ columns of $\A$ such that $\y$ has \emph{two}
distinct $k$-sparse representations in $\A$. In other words, if $\mu <
\frac{1}{2k-1}$ is not satisfied, there exist instances of
dictionaries such that \emph{no} algorithm can univocally recover some
$k$-sparse representations. In the context of Oxx, their result can be
rephrased as the following worst-case necessary condition: there
exists a dictionary $\A$ with $\mu = \frac{1}{2k-1}$ and a support
$\Qcs$, with $\vert \Qcs \vert =k$, such that Oxx selects a wrong atom
at the first iteration.

In this section, we derive a worst-case necessary condition in the
case where Oxx has selected atoms in $\Qcs$ during the first $l$
iterations. We extend Cai\&Wang's analysis and exhibit a
  scenario in which $l$ true atoms are selected, then the Oxx residual
  after $l$ iterations has two $(k-l)$-term representations. Our
result reads
\begin{theorem}[\eqref{eq:mainBound} is a worst-case necessary condition for Oxx]\label{th:NC}
There exists a dictionary $\A$ with $\mu=\frac{1}{2k-l-1}$, a support $\Qcs$ with $\vert \Qcs \vert=k$ and $\y\in\spa(\A_\Qcs)$, such that 
Oxx with $\y$ as input selects $l$ atoms in $\Qcs$ during the first $l$ iterations and a wrong atom at the $(l+1)$th iteration.
%\begin{itemize}
%\item[\emph{i)}] Oxx selects $l$ atoms in $\Qcs$ during the first $l$ iterations,
%\item[\emph{ii)}] Oxx selects a wrong atom at the $(l+1)$th iteration. 
%\end{itemize}
\end{theorem}

To reach the result, we adopt a dictionary construction
similar to Cai\&Wang's in \cite{Cai2010Stable}. Let $\M \in
\mathbb{R}^{(2k-l) \times (2k-l)}$ be the matrix with ones on the
diagonal and $-\frac{1}{2k-l-1}$ elsewhere. $\M$ will play the
  role of the Gram matrix $\M=\A^T\A$. We will exploit the eigenvalue
decomposition of $\M$ to construct the dictionary
$\A\in\mathbb{R}^{(2k-l-1)\times (2k-l)}$ with the desired properties.
Since $\M$ is symmetric, it can be expressed as
\begin{align}
\M = \U \Diag \U^T,
\end{align}
where $\U$ (resp. $\Diag$) is the unitary matrix whose
  columns are the eigenvectors (resp. the diagonal matrix of
eigenvalues) of $\M$. It is easy to check that $\M$ has only two
distinct eigenvalues: $\frac{2k-l}{2k-l-1}$ with multiplicity $2k-l-1$
and $0$ with multiplicity one; moreover, the eigenvector associated to
the null eigenvalue is equal to $\mathbf{1}_{2k-l}$. The eigenvalues
are sorted in the decreasing order so that $0$ appears in the lower
right corner of $\Lambda$.

We  define $\A\in\mathbb{R}^{(2k-l-1)\times (2k-l)}$ as
\begin{align}
\A = \Upsilon \U^T, \label{eq:defDico}
\end{align}
where $\Upsilon \in \mathbb{R}^{(2k-l-1)\times (2k-l)}$ is such that
\begin{align}
\Upsilon(i,j) = \left\{
\begin{array}{cl}
\sqrt{\frac{2k-l}{2k-l-1}} & \mbox{if $i=j$,}\\
0 & \mbox{otherwise.}
\end{array}
\right.
\end{align}
Note that $\Upsilon^T  \Upsilon=\Lambda$. Hence, $\A$ satisfies the hypotheses of Theorem \ref{th:NC} since
\begin{align}
\A^T \A = \U \Upsilon^T  \Upsilon \U^T = \U \Diag \U^T = \M,
\end{align}
and therefore 
\begin{align}
\langle \a_i, \a_j \rangle = -\frac{1}{2k-l-1} \quad \forall i \neq j.
\label{eq:prod_scal_atomes}
\end{align}
Since $\M=\A^T\A$, we have $\M\x=\mathbf{0}_{2k-l}$ if and only if
$\A\x=\mathbf{0}_{2k-l-1}$. Moreover, since $\M$ has \emph{one single}
zero eigenvalue with eigenvector $\mathbf{1}_{2k-l}$, the
  null-space of $\A$ is the one-dimensional space spanned by
  $\mathbf{1}_{2k-l}$. Therefore, any $p<2k-l$ columns of $\A$ are
linearly independent, \ie $\spark(\A)=2k-l$.

Before proceeding to the proof of Theorem \ref{th:NC}, we need to define the concept of ``reachability" of a subset $\Qc$:%the following technical lemma which provides sufficient conditions under which  OMP selects a certain atom at a given iteration: 

\begin{defi}\label{def:reachability}
A subset $\Qc$ is said to be reachable by Oxx if there exists $\y \in \spa (\A_\Qc)$ such that Oxx with $\y$ as input selects atoms in $\Qc$ during the first $\vert \Qc \vert$ iterations. 
\end{defi}

The concept of reachability was first introduced in
\cite{Soussen2012Sparsev2}. The authors showed that any subset $\Qc$
with $\vert \Qc \vert\leq \spark(\A)-2$ is reachable by OLS, see
\cite[Lemma 3]{Soussen2012Sparsev2}. On the other hand, they emphasized
that there exist dictionaries for which some subsets $\Qc$ can never
be reached by OMP, see
\cite[Example 1]{Soussen2012Sparsev2}. This scenario does however not
occur for the dictionary defined in \eqref{eq:defDico} as stated in
the next lemma:
\begin{lemma}\label{lem:SCOMP_NC}
Let $\A$ be defined as in \eqref{eq:defDico} with $l<k$. Then any subset $\Qc$ with $\vert \Qc \vert=l$ is reachable by Oxx. 
\end{lemma}

%\begin{lemma} \label{lem:SCOMP_NC} Let $\A\in\mathbb{R}^{(2k-l-1)\times (2k-l)}$ be defined as in \eqref{eq:defDico}. Assume that OMP with input $\y=\A_\Qcs \x_\Qcs$ has selected atoms in $\Qc\subset\Qcs$ during the first $l$ iterations. If
%for some $j\in \Qcs\backslash\Qc$, the following conditions hold:
%\begin{align}
%& sign(x_j) = - sign(\sum_{i\in\Qcs\backslash\Qc} x_i), \label{eq:condl3_1}\\
%& \vert x_j \vert > \vert x_i \vert\quad\mbox{$\forall\, i\in \Qcs\backslash(\Qc\cup\{j\}) $}, \label{eq:condl3_2}
%\end{align}
%then OMP selects atom $j$ at the next iteration. 
%\end{lemma}
The proof of this result is reported to Appendix \ref{app:NC}. 
To prove Theorem \ref{th:NC}, we also need the following technical lemma whose proof is reported to Appendix \ref{app:NC}:
\begin{lemma}\label{lem:Ctilde}
  Let $\A$ be defined as in \eqref{eq:defDico} with $l<k$.
    %Then, for any subset $\Qc$ with $\vert \Qc \vert=l$, we have $\forall i\notin\Qc,\,\tc_i^{\Qc}\ne\mathbf{0}_m$.} Moreover, 
    Then, for any subset $\Qc$ with $\vert \Qc \vert=l$, there
    exists a vector $\y$ having two $(k-l)$-term representations
    with disjoint supports in the projected dictionary
    $\tC_{\backslash\Qc}\triangleq \tC_{\{1,\ldots,
      2k-l\}\backslash\Qc} \in \mathbb{R}^{2k-l-1 \times 2k-2l}.$ 
\end{lemma}

We are now ready to prove Theorem \ref{th:NC}: \vspace{0.2cm}

\begin{IEEEproof}\emph{(Theorem \ref{th:NC})}
Consider the dictionary $\A$ defined in \eqref{eq:defDico}
    with $l<k$. Let $\Qc$ be a subset of cardinality $l$, arbitrarily
    chosen (say, the first $l$ atoms of the dictionary). We will
    exhibit a subset $\Qcs\supset\Qc$ for which the result of
    Theorem~\ref{th:NC} holds.

  We first apply Lemma~\ref{lem:SCOMP_NC}: there exists an
    input $\y_1\in\spa(\A_\Qc)$ for which Oxx selects all atoms in
    $\Qc$ during the first $l$ iterations. Then, we apply
    Lemma~\ref{lem:Ctilde}: there exists a vector $\y_2$ having two
    $(k-l)$-term representations in the projected dictionary
    $\tC_{\backslash\Qc}$. We will denote their respective supports by
    $\Qc_1$ and $\Qc_2$ with $\Qc_1\cap\Qc_2=\emptyset$.

  By virtue of \cite[Lemma~15]{Soussen2012Sparsev2}, Oxx with
    $\y=\y_1+\epsilon \y_2$ as input selects the same atoms (\ie
    $\Qc$) as with $\y_1$ as input during the first $l$ iterations as
    long as $\epsilon>0$ is sufficiently small. Moreover, the
    selection rule~\eqref{eq:atomselection2} indicates that the atom
    $\ta_j$ selected at iteration $l+1$ satisfies:
    \begin{align}
      j\in \ama_i  \vert \stdscal{\tc_i,\proj \y} \vert = \ama_i  \vert \stdscal{\tc_i,\y_2} \vert,\label{eq:Oxxselecstep2}
    \end{align}
    since $\proj \y = \epsilon \proj \y_2 = \epsilon \y_2$.
      Now, we set $\Qcs$ in such a way that $j\notin\Qcs$:
    \begin{align}
      \Qcs=\left\{
        \begin{array}{cl}
          \Qc\cup\Qc_1 & \mbox{if $j\in\Qc_2$},\\
          \Qc\cup\Qc_2 & \mbox{if $j\in\Qc_1$.}
        \end{array}
      \right .
      \label{eq:Qcs}
    \end{align}
    To complete the proof, it is easy to check that
      $\y=\y_1+\epsilon \y_2\in\spansub{\A_{\Qcs}}$ because
      $\y_1\in\spansub{\A_{\Qc}}$ and
      $\y_2\in\spansub{\tC_{\Qcs\backslash\Qc}}=\spansub{\tA_{\Qcs\backslash\Qc}}
      \subset\A_{\Qcs}$.
 \end{IEEEproof}

\section{Conclusions}

 The sufficient and worst-case necessary
  condition we derived for the success of Oxx after the first $l$
  iterations have been completed reads $\mu<\frac{1}{2k-l-1}$ and
  relaxes the coherence-based results by Tropp~\cite{Tropp2004Greed}
  and Cai\&Wang \cite{Cai2010Stable} corresponding to the case $l=0$.

  Our condition is obviously pessimistic since it is a worst-case
  condition for \emph{all possible} supports of cardinality $l$. In
  comparison, the conditions we elaborated
  in~\cite{Soussen2012Sparsev2} are sharper (although significantly
  more complex) and they are dedicated to a single support of size
  $l$. The latter conditions are indeed rather unpractical since they
  depend on the true support which is unknown. In practice, they shall
  be evaluated for all possible pairs of complete/partial supports of
  dimension $k$ and $l$, and each evaluation requires a pseudo-inverse
  computation. A compromise between the pessimistic coherence condition and
  those elaborated in~\cite{Soussen2012Sparsev2} would be to adapt our
  mutual coherence results to the cumulative
  coherence~\cite{Tropp2004Greed}, and the weak ERC
  condition~\cite{Tropp2004Greed,Dossal2005b,Gribonval2008} (also
  referred to as the Neumann ERC in~\cite{Lorenz2011}). The latter
  conditions are intermediate conditions at iteration 0 between the
  mutual coherence condition $\mu<1/(2k-1)$ and Tropp's ERC.  Their
  computation remains simple as only inner products between the
  dictionary atoms are involved.  It would therefore be definitely interesting to study how this type of condition evolve when Oxx has recovered $l$ atoms of the support. This is part of our future work.

In this paper, we did also not investigate the case where the observed
vector $\y$ is corrupted by some additive noise. This problem has been
addressed in different contributions of the recent literature, see \eg
\cite{Donoho2006Stable,Cai2011Orthogonal}, and is interesting on its
own. The extension of the proposed partial condition to noisy settings
is part of our ongoing work.

% In this paper, we derive a new condition, based on the coherence of the dictionary, for the success of Oxx when $l$ atoms of the support have been selected during the first $l$ iterations. Our result reads $\mu<\frac{1}{2k-l-1}$ and generalizes previous coherence-based results by Tropp \cite{Tropp2004Greed} and Cai\&Wang \cite{Cai2010Stable} to the case where Oxx has some partial knowledge of the support. In particular, we show that the proposed condition is not only sufficient but also worst-case necessary in the following sense: there exists a dictionary with $\mu=\frac{1}{2k-l-1}$ such that Oxx will select atoms of the support during the first $l$ iterations and then selected a wrong atom. 

\appendices
\section{Proof of the results of section \ref{sec:SCOMP}}

This section contains the proofs of Propositions~\ref{UBERC} and
\ref{prop:linkRICmu} together with some useful technical lemmas.
%We first need to 
%define the coherence\footnote{The
%  standard definition of the coherence usually assumes that the
%  columns of the dictionary are normalized. We thus use the term
%  "coherence" with a slight abuse of language here since the columns
%  of $\proj \A$ are not necessarily normalized to 1.} of the projected
%dictionary as follows:
% and drive a link between the latter and the upper/lower bounds defined in Definition \ref{def:GRIC}:
%
%\begin{defi}
%\begin{align}
%\mu^{OMP}_l= \max_{\vert \Qc \vert=l}\max_{i \neq j}  \vert \langle \ta_i, \ta_j  \rangle\vert. 
%\end{align}
%\end{defi}
%When $l=0$, one recovers the standard definition of the coherence of the dictionary.
% The next lemma relates the coherence of the projected dictionary to the isometry bounds of the projected dictionary $\ud_{2,l}$ and $\ld_{2,l}$:
\begin{lemma}\label{len:lemma1} Assume $\A$ satisfies the  P-RIP($\ld_{2,l}$,$\ud_{2,l}$) and let
\begin{align}
\mu^{OMP}_l\triangleq \max_{\vert \Qc \vert=l}\max_{i \neq j}  \vert \langle \ta_i^\Qc, \ta_j^\Qc  \rangle\vert. 
\end{align}
Then, we have
\begin{align}
%\mu^{OMP}_l \leq \frac{\ub(2,l)-\lb(2,l)}{2}.
\mu^{OMP}_l \leq \frac{\ud_{2,l}+\ld_{2,l}}{2}.
\end{align}
\end{lemma}
\begin{IEEEproof}
By definition of $\ud_{2,l}$ and $\ld_{2,l}$ we must have for all $\Qc, \Qc'$ with $\vert \Qc \vert=l$, $\vert \Qc' \vert=2$ and $\Qc'\cap\Qc=\emptyset$:
\begin{align}
1+\ud_{2,l}\geq \lambda_{max}(\tilde{\A}_{\Qc'}^T \tilde{\A}_{\Qc'}),\label{eq:lbub}\\
1-\ld_{2,l} \leq \lambda_{min}(\tilde{\A}_{\Qc'}^T \tilde{\A}_{\Qc'}),\label{eq:ublb}
\end{align}
where $\lambda_{max}(\M)$ (resp. $\lambda_{min}(\M)$) denotes the
largest (resp. smallest) eigenvalue of $\M$. Moreover, if
${\Qc'}=\{i,j\}$, it is easy to check that the eigenvalues of
$\tilde{\A}_{\Qc'}^T \tilde{\A}_{\Qc'}$ can be expressed as
\begin{align}
\lambda_{}(\tilde{\A}_{\Qc'}^T \tilde{\A}_{\Qc'})=\frac{\| \tilde{\a}_i\|^2+\| \tilde{\a}_j\|^2 \pm \Delta}{2},\nonumber
%\lambda_{min}(\tilde{\A}_{\Qc'}^T \tilde{\A}_{\Qc'})=\frac{\| \tilde{\a}_i\|^2+\| \tilde{\a}_j\|^2 - \Delta}{2},\nonumber
\end{align}
where
\begin{align}
\Delta &=\sqrt{(\| \tilde{\a}_i\|^2+\| \tilde{\a}_j\|^2)^2+4(
    \langle \tilde{\a}_i, \tilde{\a}_j \rangle^2-\| \tilde{\a}_i\|^2
    \: \| \tilde{\a}_j\|^2) }  \\
&=\sqrt{(\| \tilde{\a}_i\|^2-\| \tilde{\a}_j\|^2)^2+4 \langle \tilde{\a}_i, \tilde{\a}_j \rangle^2 }. 
\end{align}
Hence
\begin{align}
\lambda_{max}(\tilde{\A}_{\Qc'}^T \tilde{\A}_{\Qc'})-\lambda_{min}(\tilde{\A}_{\Qc'}^T \tilde{\A}_{\Qc'})
&= \Delta \geq 2 \vert \langle \tilde{\a}_i, \tilde{\a}_j \rangle \vert. \nonumber
\end{align}
Using \eqref{eq:lbub}-\eqref{eq:ublb}, we thus obtain $\forall i, j \notin \Qc$:
\begin{align}
\ud_{2,l}+\ld_{2,l} \geq 2 \vert \langle \tilde{\a}_i, \tilde{\a}_j \rangle \vert.\label{eq:ineqinterqqq}
\end{align}
Now, this inequality also holds if $i\in \Qc$ or $j\in \Qc$ since the
right hand-side of \eqref{eq:ineqinterqqq} is then equal to zero.  The
result then follows from the definition of $\mu_l^{OMP}$.
\end{IEEEproof}

\begin{lemma} \label{len:lemma2} Let $\vert \Qc \vert$=l and $\Qc' \cap \Qc'' =\emptyset$, then $\forall \u \in \mathbb{R}^{\vert\Qc''\vert}$,
\begin{align}
\| \tA_{\Qc'}^T  \tA_{\Qc''} \u  \|\leq  \mu^{OMP}_l \sqrt{\vert {\Qc'} \vert \vert {\Qc''} \vert} \, \|\u\|.
\end{align}
\end{lemma}

 \begin{IEEEproof}
 We have
 \begin{align}
\| \tilde{\A}_{\Qc'}^T \tilde{\A}_{\Qc''} \u  \| 
&= \sqrt{\sum_{i\in {\Qc'}} \langle \tilde{\a}_i, \tilde{\A}_{\Qc''} \u \rangle^2}\\
&=\sqrt{\sum_{i\in {\Qc'}} 
\bigl (\sum_{j\in {\Qc''}} u_j\, \langle \tilde{\a}_i, \tilde{\a}_j \rangle
\bigr)^2}\\
&\leq \sqrt{\sum_{i\in {\Qc'}} \bigl(\sum_{j\in {\Qc''}} \vert u_j\vert \, \vert\langle \tilde{\a}_i, \tilde{\a}_j \rangle\vert\bigr)^2}\\
&\leq  \mu^{OMP}_l \sqrt{\vert {\Qc'} \vert}\, \| \u \|_1\\
&\leq  \mu^{OMP}_l \sqrt{\vert {\Qc'} \vert \vert {\Qc''} \vert}\, \| \u \|.
\end{align}
\end{IEEEproof}

Using Lemmas~\ref{len:lemma1} and~\ref{len:lemma2}, we can now prove Propositions \ref{UBERC} and \ref{prop:linkRICmu}:\vspace{0.2cm}

\begin{IEEEproof}
\emph{(Proposition \ref{UBERC})} $\forall\, i\notin \Qcs$, the following inequalities hold:
\begin{align}
\| \tA_{\Qc^\star \backslash \Qc}^\dag \tilde{\a}_{i} \|_1 
&\leq \sqrt{k-l} \,\| \tA_{\Qc^\star \backslash \Qc}^\dag \tilde{\a}_{i} \|_2,\\
&\leq \frac{\sqrt{k-l}}{1-\ld_{k-l,l}} \| \tA_{\Qc^\star \backslash \Qc}^T \tilde{\a}_{i} \|_2,\\
&\leq \frac{k-l}{1-\ld_{k-l,l}} \mu^{OMP}_l,\\
&\leq \frac{k-l}{1-\ld_{k-l,l}} \frac{\ud_{2,l}+\ld_{2,l}}{2},
\end{align}
where the first inequality follows from the equivalence of norms; the second from RIC properties (see \cite[Proposition 3.1]{Needel_ACMels08}); the third from Lemma \ref{len:lemma2} and the fourth from Lemma \ref{len:lemma1}. 
\end{IEEEproof}\vspace{0.2cm}

\begin{IEEEproof}
\emph{(Proposition \ref{prop:linkRICmu})} 
First, notice that $\A$ satisfies the P-RIP($\ld_{q,0}$,$\ud_{q,0}$) $\forall \, q$ with
\begin{align}
\ud_{q,0}=\ld_{q,0}&= (q-1)\mu,
%\ud_{q,0}&= (q-1)\mu,
\end{align}
%\begin{align}
%1-(q-1)\mu \leq \lb(q,0) \leq \ub(q,0)\leq 1+(q+1)\mu,\quad \forall q
%\end{align}
see \eg \cite[Lemma 2.3]{Tropp2004Greed}. 
Hence, \eqref{eq:relRICmu1} is a consequence of the following inequalities:
\begin{align}
 \| \proj \A_{\Qc '} \xQp \|^2 \leq  \| \AQp \xQp \|^2 \leq (1+\ud_{q,0}) \|\xQp \|^2.
\end{align}
Lower bound \eqref{eq:relRICmu2} may derived by noticing that
\begin{align}
\| \proj \AQp \xQp \|^2 &= \| \AQp \xQp \|^2 -\| \mathbf{P}_\Qc\AQp \xQp \|^2,
\end{align}
and 
\begin{align}
\| \AQp \xQp \|^2&\geq (1-\ld_{q,0}) \|\xQp \|^2,\\
\| \mathbf{P}_\Qc\AQp \xQp \|^2
&=\|(\A_\Qc^\dag)^T \A_\Qc^T \A_{\Qc'} \x_{\Qc'}\|^2\\
&\leq \frac{\|\A_\Qc^T \A_{\Qc'} \x_{\Qc'}\|^2}{1-\ld_{l,0}}\label{eq:ineqPRIP},\\
&\leq \frac{\mu^2 l q \, \| \x_{\Qc'}\|^2}{1-\ld_{l,0}},\label{eq:conlemma2}
\end{align}
where inequality \eqref{eq:ineqPRIP} follows from standard
  relationships between the RIC properties of $\A$ and transforms of
  $\A$,  and $1-\ld_{l,0}\geq 0$ is a consequence of hypothesis
$\mu<1/(l-1)$~\cite[Lemma~2.3]{Tropp2004Greed}; \eqref{eq:conlemma2}
is a consequence of Lemma \ref{len:lemma2}.
\end{IEEEproof}

\section{Proof of the results of section \ref{sec:SCOLS}}\label{annex:SCOLS}

\begin{IEEEproof} \emph{(Lemma \ref{prop:bound_mutilde})}
The proof is recursive. Obviously, the result holds for 
$l=0$ since $\mu^{OLS}_0=\mu$.

Let $\Qc$ with $\vert\Qc\vert=l\geq1$ and consider $\Rc$ such that
$\Qc=\Rc\cup{\{i\}}$ with $\vert\Rc\vert=l-1$. According
to~\cite[Lemma~5]{Soussen2012Sparsev2},
if $j\notin\Qc$, we have the orthogonal decomposition
\begin{align}
  \tilde{\b}_{j}^{\Rc} = \eta_{j}\tilde{\b}_{j}^{\Qc}+
  \langle \tilde{\b}_{j}^{\Rc} \,,\, \tilde{\b}_{i}^{\Rc} \rangle
  \,\tilde{\b}_{i}^{\Rc}.
  \label{eq:orthog_decomp}
\end{align}
 Moreover, assumption $\mu<1/l$ implies that
  $\A_{\Qc\cup\{j\}}$, $\A_{\Rc\cup\{j\}}$ and $\A_{\Rc\cup\{i\}}$ are
  full column rank as families of at most $l+1$
  atoms~\cite[Lemma~2.3]{Tropp2004Greed} which in turn implies that
  $\tilde{\a}_{j}^{\Qc}$, $\tilde{\a}_{j}^{\Rc}$ and
  $\tilde{\a}_{i}^{\Rc}$ are
  nonzero~\cite[Corollary~3]{Soussen2012Sparsev2}. Therefore,
  $\|\tilde{\b}_{j}^{\Qc}\|$, $\|\tilde{\b}_{j}^{\Rc}\|$ and
  $\|\tilde{\b}_{i}^{\Rc}\|$ are all of unit norm, and 
then~\eqref{eq:orthog_decomp} yields $\eta_j=\pm \sqrt{1-\langle
  \tilde{\b}_{j}^{\Rc} \,,\, \tilde{\b}_{i}^{\Rc} \rangle^2}$.
If $j$ and $j'\notin\Qc$, it follows that 
\begin{align}
  \langle \tilde{\b}_{j}^{\Qc}\,,\,\tilde{\b}_{j'}^{\Qc}\rangle&=
  \frac{
  \langle \tilde{\b}_{j}^{\Rc}\,,\,\tilde{\b}_{j'}^{\Rc} \rangle -
  \langle \tilde{\b}_{j}^{\Rc} \,,\, \tilde{\b}_{i}^{\Rc} \rangle
  \langle \tilde{\b}_{j'}^{\Rc} \,,\, \tilde{\b}_{i}^{\Rc} \rangle}{
  \eta_{j}\eta_{j'}}.
\end{align}
Majorizing the inner products $\vert \langle \tilde{\b}_{j}^{\Rc}
\,,\, \tilde{\b}_{i}^{\Rc} \rangle\vert$ by $\mu^{OLS}_{l-1}$
and using~\eqref{eq:bound_mutilde}, we get:
\begin{align}
  \vert\langle\tilde{\b}_{j}^{\Qc}\,,\,\tilde{\b}_{j'}^{\Qc}\rangle
  \vert&\leqslant \frac{\mu^{OLS}_{l-1}+(\mu^{OLS}_{l-1})^2}{
  1-(\mu^{OLS}_{l-1})^2}\\
&= \frac{\mu^{OLS}_{l-1}}{1-\mu^{OLS}_{l-1}}\\
&\leqslant \frac{\mu}{1-(l-1)\mu-\mu}=\frac{\mu}{1-l\mu}
\end{align}
leading to~\eqref{eq:bound_mutilde}.
\end{IEEEproof}

\section{Proof of the results of section \ref{sec:NC}}\label{app:NC}

In this appendix, we provide a proof of Lemma \ref{lem:SCOMP_NC}.
We use the notation $\Rc$ instead of $\Qc$ to denote the
  current support. This change of notation is done to avoid confusion:
  in the rest of the paper, we have $\vert\Qc\vert=l$ whereas in this
  appendix, the support cardinality may differ from $l$.

We first need to prove the following technical lemma:
\begin{lemma} \label{lem:symproblem} Let $\A$ be defined as in
  \eqref{eq:defDico}. Then, we have for all $\Rc$ with $\vert
  \Rc \vert <2k-l$ and $i, j\notin \Rc$, $i\neq j$:
%Then, we have $\forall\, \Qc$ with $\vert \Qc \vert = l<2k-l$ and $h,i, j\notin \Qc$:
%\begin{align}
%\stdscal{\ta_h,\ta_i} &= \stdscal{\ta_h,\ta_j}<0\quad  \mbox{$i\neq h, j\neq h $},\label{lem:c1}\\
%\| \ta_i \| &= \| \ta_j \|.
%\end{align}
\begin{align}
\stdscal{\ta_i^{\Rc},\ta_j^{\Rc}}&=-\mu-\mu^2\mathbf{1}_{\vert\Rc\vert}^T (\A_\Rc^T \A_\Rc)^{-1} \mathbf{1}_{\vert\Rc\vert},\label{lem:c1}\\
\| \ta_i^{\Rc}\|^2&= 1-\mu^2\mathbf{1}_{\vert\Rc\vert}^T (\A_\Rc^T \A_\Rc)^{-1} \mathbf{1}_{\vert\Rc\vert}.\label{lem:c1b}
\end{align}
\end{lemma}

\begin{IEEEproof} First recall that $\spark(\A)=2k-l$ (see section
  \ref{sec:NC}). Therefore, $\A_\Rc$ is full rank when $\vert
  \Rc\vert<2k-l$ and $\ta_i^{\Rc}$ reads
\begin{align}
\ta_i^{\Rc} = 
\mathbf{P}_\Rc^\bot 
\a_i= \a_i - \mathbf{P}_{\Rc}\a_i = \a_i - \A_{\Rc} (\A_{\Rc}^T \A_{\Rc})^{-1} \A_{\Rc}^T \a_i.
\end{align}
Using this expression, we have
\begin{align}
\stdscal{\ta_i^{\Rc},\ta_j^{\Rc}}&=\stdscal{\a_i,\a_j}-\a_i^T\A_{\Rc} (\A_{\Rc}^T \A_{\Rc})^{-1} \A_{\Rc}^T \a_j,\\
\| \ta_i^{\Rc}\|^2&= 1- \a_i^T\A_{\Rc} (\A_{\Rc}^T \A_{\Rc})^{-1} \A_{\Rc}^T \a_i.
\end{align}
Taking into account that the inner product between any pair of atoms is equal to $-\mu$ by definition of $\M=\A^T\A$, we obtain the result. %these expressions become
%\begin{align}
%\stdscal{\ta_i,\ta_j}&=-\mu-\mu^2\mathbf{1}_{l}^T (\A_\Qc^T \A_\Qc)^{-1} \mathbf{1}_{l},\\
%\| \ta_i\|^2&= 1-\mu^2\mathbf{1}_{l}^T (\A_\Qc^T \A_\Qc)^{-1} \mathbf{1}_{l}.
%\end{align}

\end{IEEEproof}

\begin{IEEEproof} (\emph{Lemma \ref{lem:SCOMP_NC}})
  We prove a result slightly more general than the statement of Lemma
  \ref{lem:SCOMP_NC}: for the dictionary defined as in
  \eqref{eq:defDico}, any subset $\Rc$ with
    $p\triangleq\vert\Rc \vert\leq 2k-l-2$ can be reached by Oxx.
  Lemma \ref{lem:SCOMP_NC} corresponds to the case $p=l$
  ($p\leq 2k-l-2$ is always satisfied as long as $l<k$).

  The result is true for OLS by virtue of
  \cite[Lemma~3]{Soussen2012Sparsev2} which states that any subset
  $\Rc$ of an \emph{arbitrary} dictionary $\A$ is reachable as long as
  $\vert \Rc\vert\leq \spark(\A)-2$.  In particular, the latter
  condition is verified by the dictionary $\A$ and the subset $\Rc$
  considered here since $\spark(\A)=2k-l$ and $\vert\Rc \vert\leq
  2k-l-2$ by hypothesis.

  We prove hereafter that the result is also true for OMP. Without
  loss of generality, we assume that the elements of $\Rc$ correspond
  to the first $p$ atoms of $\A$ (the analysis
    performed hereafter remains valid for any other support $\Rc$ of
    cardinality $p$ since the content of the Gram matrix
    $\A_\Rc^T\A_\Rc$ is constant whatever the support $\Rc$:
    see~\eqref{eq:prod_scal_atomes}).
  % From the definition of
  % reachability (see Definition \ref{def:reachability}) we want
  % therefore to show that there exists $\y\in\spa\{\a_1, \ldots,
  % \a_q\}$ such that OMP with $\y$ as input selects atoms in
  % $\Qc=\{1,\ldots,q\}$ during the first $q$ iterations.
  %
  % 
  For arbitrary values of $\epsilon_2,\ldots, \epsilon_{p}>0$,
  we define the following recursive construction:
  \begin{itemize}
    \item $\y_1=\a_1$,
    \item $\y_{p+1}=\y_{p}+\epsilon_{p+1}\a_{p+1}$ 
  \end{itemize}
  ($\y_{p+1}$ implicitly depends on
  $\epsilon_2,\ldots,\epsilon_{p+1}$).  We show by recursion
  that for all $p\in\{1,\ldots,2k-l-2\}$, there exist
  $\epsilon_2,\ldots,\epsilon_{p}>0$ such that OMP with the dictionary
  defined as in~\eqref{eq:defDico} and $\y_{p}$ as input successively
  selects $\a_1,\ldots,\a_{p}$ during the first $p$ iterations (in
  particular, the selection rule~\eqref{eq:atomselection2} always
  yields a unique maximum).

  The statement is obviously true for $\y_1=\a_1$. Assume that it is
  true for $\y_{p}$ ($p<2k-l-2$) with some $\epsilon_2,\ldots,
  \epsilon_{p}>0$ (these parameters will remain fixed in the
  following). According to \cite[Lemma 15]{Soussen2012Sparsev2}, there
  exists $\epsilon_{p+1}>0$ such that OMP with
  $\y_{p+1}=\y_{p}+\epsilon_{p+1}\a_{p+1}$ as input selects the same
  atoms as with $\y_{p}$ during the first $p$ iterations, \ie
  $\a_1,\ldots,\a_{p}$ are successively chosen. At iteration $p$, the
  current active set reads $\Rc=\{1,\ldots,p\}$ and the corresponding
  residual takes the form
  \begin{align}
    \r_{\Rc}= \epsilon_{p+1} \ta_{p+1}^{\Rc}.
  \end{align}
  Thus, $\a_{p+1}$ is chosen at iteration $p+1$ if and only if
  \begin{align}\label{eq:goodselectionOMP}
    \vert\stdscal{\ta_i^{\Rc},\ta_{p+1}^{\Rc}}\vert< 
\| \ta_{p+1}^{\Rc} \|^2 \qquad \forall \, i\neq p+1. 
  \end{align}

  Now, $\vert \Rc\vert=p<2k-l$ by hypothesis, then Lemma
  \ref{lem:symproblem} applies. Using \eqref{lem:c1}-\eqref{lem:c1b},
  it is easy to see that \eqref{eq:goodselectionOMP} is equivalent to
  \begin{align}\label{eq:cnsgoodselectionOMP}
    \mu + 2 \mu^2 \mathbf{1}_{p}^T (\A_{\Rc}^T \A_{\Rc})^{-1} \mathbf{1}_{p} < 1.
  \end{align}
  Since $\mu=\frac{1}{2k-l-1}<\frac{1}{p+1}<\frac{1}{p-1}$, we have
  $(1-(p-1)\mu)>0$. Then, \cite[Lemma~2.3]{Tropp2004Greed} and
    $\|\mathbf{1}_{p}\|^2=p$ yield:
  \begin{align}
    \mathbf{1}_{p}^T (\A_{\Rc}^T \A_{\Rc})^{-1} \mathbf{1}_{p}\leq \frac{p}{1-(p-1)\mu}. 
  \end{align}
Using the majoration $\mu<1/(p+1)$, it follows that:
  \begin{align}
    \mu + 2 \mu^2 \mathbf{1}_{p}^T (\A_{\Rc}^T \A_{\Rc})^{-1}
    \mathbf{1}_{p} &\leqslant
    \mu\left (1+  \frac{2 \mu p}{1-(p-1)\mu}\right )\\
    &= \mu\left (\frac{1+(p+1)\mu}{1-(p-1)\mu}\right )\\
    &< \frac{1}{p+1}\left (\frac{2}{1-\frac{p-1}{p+1}}\right )=1
  \end{align}
which proves that the
    condition~\eqref{eq:cnsgoodselectionOMP}, and
    then~\eqref{eq:goodselectionOMP} is met.
  % \begin{align}
  %   \mu<\frac{1}{p+1}. 
  % \end{align}
  % or equivalently
  % \begin{align}
  %   p<2k-l-2. 
  % \end{align}
OMP therefore recovers the subset
    $\Rc\cup\{p+1\}=\stdacc{1,\ldots,p+1}$.

\end{IEEEproof}

\begin{IEEEproof}\emph{(Lemma~\ref{lem:Ctilde})}
 % \addCS{Since $\spark(\A)=2k-l\geqslant  2(l+1)-l=l+2$, $\A_{\Qc\cup\{i\}}$ is full rank for $i\notin\Qc$,  so $\tc_i^{\Qc}\ne\textbf{0}$ according  to~\cite[Corollary~3]{Soussen2012Sparsev2}.}
% 
Using Lemma \ref{lem:symproblem},
    we notice that $\tC_{\backslash\Qc}=\beta\tA_{\backslash\Qc}$ for
    some $\beta>0$ since $\| \ta_i \|$ does not depend on $i$ and
    $\tc_i\ne\textbf{0}$. Defining $\v\triangleq\mathbf{1}_{2k-2l}$,
    we obtain
  \begin{align}
    \tC_{\backslash\Qc}\v
    &= \beta \tA_{\backslash\Qc}\v\\
    & =\beta\tA \mathbf{1}_{2k-l}= \beta \proj\A\mathbf{1}_{2k-l} = \mathbf{0}_{2k-l-1}, \label{eq:detailsNC1}
  \end{align}
  since $ \mathbf{1}_{2k-l}$ belongs to the null-space of
    $\A$.

Let us partition the elements of $\v=\mathbf{1}_{2k-l}$ into
    two subsets $\Qc_1\cup\Qc_2$ with $\Qc_1\cap \Qc_2=\emptyset$ and
    $\vert \Qc_1\vert=\vert \Qc_2\vert=k-l$, and define
    $\y\triangleq\tC_{\Qc_1\backslash\Qc}\mathbf{1}_{k-l}$. According
    to~\eqref{eq:detailsNC1}, $\y$ rereads
    $-\tC_{\Qc_2\backslash\Qc}\mathbf{1}_{k-l}$, therefore $\y$ has
    two $(k-l)$-sparse representations with disjoint supports in
    $\tC_{\backslash\Qc}$. 

\end{IEEEproof}

\bibliographystyle{IEEEbib}
%\bibliography{../../../References/bibliography_new.bib}
\bibliography{group-15302}

\end{document}